%% file: main.tex
\documentclass[sigconf, nonacm]{acmart} 

\AtBeginDocument{%
  } 

\setcopyright{none}
\copyrightyear{2025}
\acmYear{2025}
\acmDOI{}
\acmConference[CIKM '25 Workshop LASS]{ACM International Conference on Information and Knowledge Management}{November 10--14, 2025}{Seoul, Korea}

\usepackage{enumitem} 

\begin{document}


\title{Simulating Misinformation Propagation in Social Networks using Large Language Models}

\author{Raj Gaurav Maurya}
\authornote{Corresponding author}
\email{rajg.maurya@tum.de}
\orcid{0009-0001-5581-4840}
\affiliation{%
  \institution{Technische Universität}
  \city{München}
  \country{Germany}
}

\author{Vaibhav Shukla}
\orcid{0009-0005-1345-1710}
\affiliation{%
  \institution{Friedrich-Alexander-Universität}
  \city{Erlangen-Nürnberg}
  \country{Germany}
}

\author{Raj Abhijit Dandekar}
\author{Rajat Dandekar}
\author{Sreedath Panat}
\affiliation{%
  \institution{Vizuara AI Labs}
  \city{Pune}
  \country{India}
}
\begin{abstract}
Misinformation on social media thrives on surprise, emotion, and identity-driven reasoning, often amplified through human cognitive biases. 
To investigate these mechanisms, we model large language model (LLM) personas as synthetic agents that mimic user-level biases, ideological alignments, and trust heuristics. Within this setup, we introduce an auditor–node framework to simulate and analyze how misinformation evolves as it circulates through networks of such agents. News articles are propagated across networks of persona-conditioned LLM nodes, each rewriting received content. A question–answering (QA)-based auditor then measures factual fidelity at every step, offering interpretable, claim-level tracking of misinformation drift. 
We formalize a misinformation index (MI) and a misinformation propagation rate (MPR) to quantify factual degradation across homogeneous and heterogeneous branches of up to 30 sequential rewrites. Experiments with 21 personas across 10 domains reveal that identity- and ideology-based personas (e.g., religious leaders, lifestyle influencers, politically aligned individuals) act as misinformation accelerators, especially in politics, marketing, and technology. By contrast, expert-driven personas (e.g., medical professionals, investigative journalists) preserve factual stability. Controlled-random branch simulations further show that once early distortions emerge, heterogeneous persona interactions rapidly escalate misinformation to propaganda-level distortion. Our taxonomy of misinformation severity—spanning factual errors, lies, and propaganda—connects observed drift to established theories in misinformation studies. These findings demonstrate the dual role of LLMs as both proxies for human-like biases and as auditors capable of tracing information fidelity. The proposed framework provides an interpretable, empirically grounded approach for studying, simulating, and mitigating misinformation diffusion in digital ecosystems.
\end{abstract}

\ccsdesc[500]{Computing methodologies~Natural language processing}
\ccsdesc[500]{Computing methodologies~Multi-agent systems}
\ccsdesc[300]{Human-centered computing~Collaborative and social computing theory, concepts and paradigms}
\ccsdesc[300]{Information systems~Data mining}
\ccsdesc[300]{Security and privacy~Human and societal aspects of security and privacy}
\ccsdesc[100]{Applied computing~Psychology}
\ccsdesc[100]{Applied computing~Sociology}

\keywords{large language models, social simulation, social networks, misinformation, fake news, propaganda, persona, agents, QA, auditor, node}

\begin{teaserfigure}
\centering
  \includegraphics[width=\textwidth]{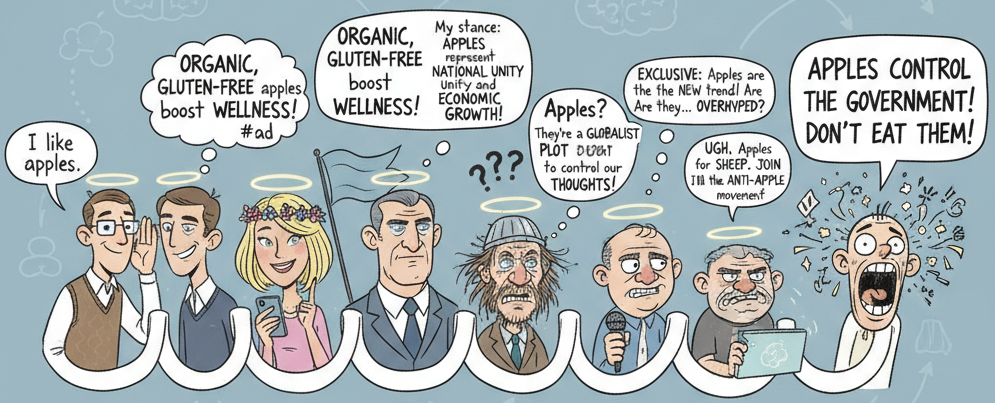}
  \caption{How agents propagate misinformation? [Generated using Google AI Studio's Nano Banana]}
  \Description{Teaser figure}
  \label{fig:teaser}
\end{teaserfigure}

\maketitle

\section{Introduction}
Misinformation on social media undermines public trust and information quality. \citet{vosoughi_spread_2018} analyzed 126,000 true and false news stories on X\footnote{formerly, Twitter: \url{https://x.com/}} and found that false news spreads faster and farther, largely due to human behavior rather than bots. False stories are more surprising and emotional, making them more shareable even when spread by users with fewer followers. The typology by \citet{tandoc_defining_2018} categorizes fake news into six types: satire, parody, fabrication, manipulation, advertising, and propaganda; highlighting how it mimics legitimate formats and is amplified on social media through blurred source boundaries and audience validation. Human judgment, social motives, and sharing dynamics thus remain central to its diffusion.

Individual interpretation of information is shaped by beliefs, experiences, and identities \citep{pennycook_fighting_2019}, which motivates using Large Language Models (LLMs) as synthetic agents. Trained on diverse data, LLMs capture linguistic styles, emotional tones, biases, and perspectives, enabling persona-driven simulations of real users. Similar to dynamics observed in \citet{cinelli_covid-19_2020}, LLM-based personas can be conditioned to respond differently based on source credibility and ideological alignment, mirroring human trust biases. Inspired by weakly supervised veracity classification \citep{leite_weakly_2023}, our framework incorporates weighted credibility mechanisms, while \citet{chen_can_2024} underscores the dual role of LLMs in both generating and detecting misinformation, informing safeguards against hallucination risks. As argued by \citet{thapa_large_2025}, these capabilities make LLMs promising tools in computational social science for modeling ideological bias and social amplification.

Recent reviews \citep[e.g., ][]{jerez-villota_understanding_2025} show increasing use of digital twins and LLM-based simulations, which more authentically integrate cognitive and network interactions. Evidence from \citet{he_digital_2024} and \citet{lin_human_2024} on human digital twins demonstrates architectures for bidirectional user modeling, inspiring LLM-based agents that embed cognitive and social data flows. Building on \citet{liu_skepticism_2024}, we extend personality-conditioned LLMs with memory and reflection to capture dynamic shifts in attitudes and propagation trends. Grounded in theories of cognition and social simulation \citep{conte_manifesto_2012, castelfranchi_theory_2001}, our framework enables more realistic modeling of echo chambers and peer reinforcement than traditional approaches.

Misinformation is further shaped by motivated cognition. \citet{dash_persona-assigned_2025} show that persona-conditioned LLMs replicate identity-driven reasoning, reducing accuracy by up to 9\% and aligning endorsement with political identity by up to 90\%, even resisting prompt-based debiasing. These findings align with prior work on persistent misinformation effects \citep{lewandowsky_misinformation_2012}. Supporting evidence from \citet{ward_evaluating_2024, pratelli_evaluating_2025} confirms that persona-conditioned LLMs embed cognitive tendencies and ideological biases, justifying their role as synthetic agents in misinformation simulations.

Further, \citet{mittelstadt_large_2024} show that LLMs achieve human-level performance in situational judgment tests, demonstrating nuanced social reasoning. These capabilities justify their use as persona proxies in misinformation spread, where cognition, trust, identity, and network interactions co-construct propagation. By incorporating persona effects, source credibility, and social cues, simulations capture the psychological and social drivers of misinformation more realistically.

Finally, LLMs also provide auditing mechanisms. Building on \textsc{QAFactEval} \citep{fabbri_qafacteval_2022}, LLM auditors can generate targeted questions and assign binary presence scores to track information retention. This interpretable and scalable evaluation approach, validated by \citet{aher_using_2023}, rigorously measures fidelity of information during persona-driven transformations. Thus, persona-conditioned LLMs, combined with credibility weighting and auditing, offer a powerful, empirically grounded framework for simulating and mitigating misinformation spread.

In this work, we introduce an ``auditor—node" framework to rigorously quantify how factual information is preserved or degraded as news articles propagate through networks of LLMs. First, we introduce our system design, the propagation model, and evaluation metrics. Then, we present and analyze the results of our experiments done on two different setups, before discussing the general implications and final conclusions: a summary of our contribution and future outlook is also provided.

\section{Methodology}
\subsection{Simulation Framework}
\label{sec:experimental-setup}
\begin{figure}[htbp]
    \centering
    \includegraphics[width=\linewidth]{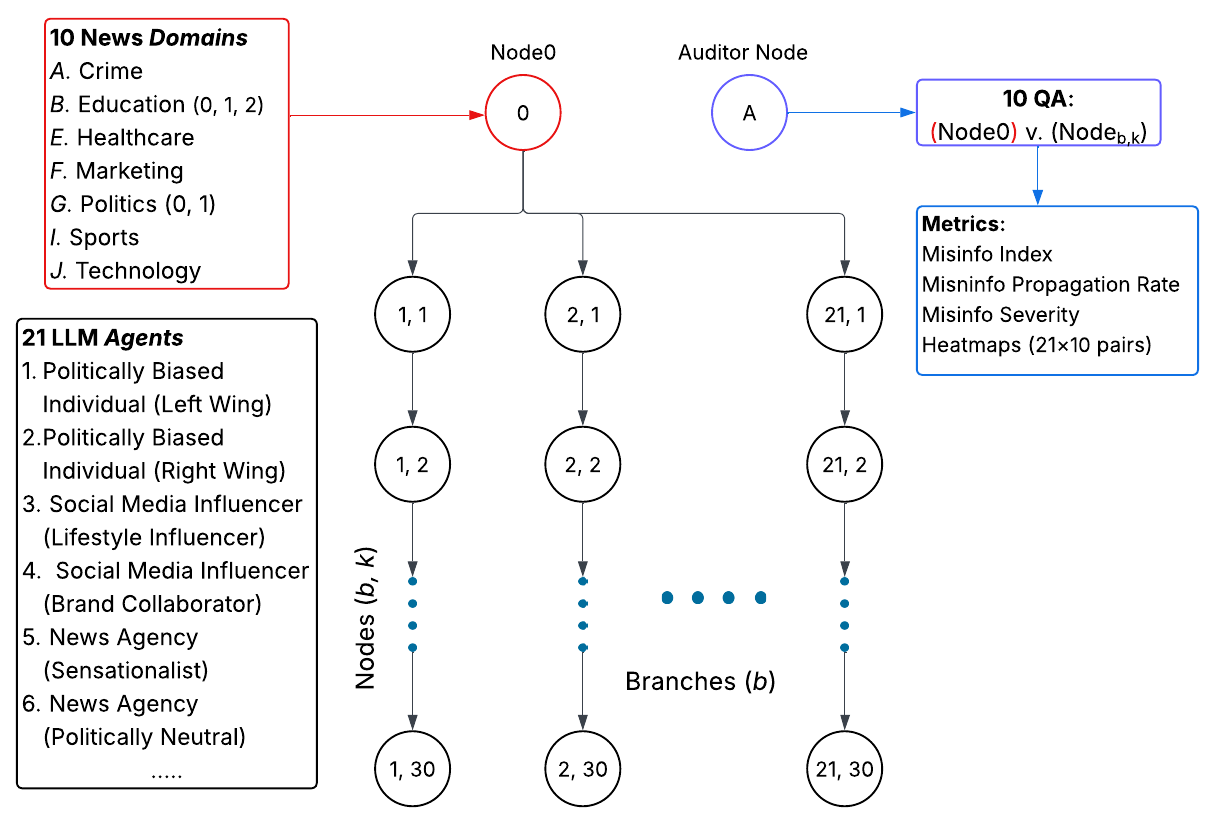}
    \caption{Diagram of system architecture for misinformation propagation, illustrating the flow of information across persona-conditioned LLM nodes, the auditor’s intervention, and data recording modules. This design allows tracing factual drift step-by-step across heterogeneous or homogeneous branches.}
    \Description{A diagram}
    \label{fig:misinformation_flow}
\end{figure}

As illustrated in Fig. \ref{fig:misinformation_flow}, our simulation consists of a network of 21 branches, each containing 30 nodes. Each node ($b, k$) corresponds to a LLM \textit{agent} (conditioned with one out of 21 distinct personas) that rewrites a news article (\textit{domain}) incoming from the node ($b, k$-$1$) above it. The original article (one of 10) is fed throughout the network by the neutral node ($b, 0$ or \texttt{Node0}), while an external \textit{auditor} (at \texttt{NodeA}) evaluates factual fidelity at every step by ``asking" 10 questions and comparing with original answers.

\subsubsection{Propagation Model}
We simulate information diffusion by propagating each of the 10 domains (one at a time) through all the 21 branches, node by node (1 to 30). After each rewrite of the news, the LLM passes a copy of the output to the next node/LLM in the branch for it to be rewritten again and another copy to the auditor for it to be evaluated against the original news (i.e., output of neutral agent at \texttt{Node0}). 
Formally, let the original article be denoted by \(S\). The auditor constructs \(m=10\) factual questions \(Q\),
\begin{equation}
Q = \{ q_j \}_{j=1}^m, \qquad G = \{ g_j \}_{j=1}^m,
\end{equation}
with corresponding correct answers \(G\).  
The article propagates through \(B\) independent branches, each of fixed length \(K=30\). In branch \(b\), node \(k\) receives the previous article \(X_{b,k-1}\) and outputs a rewritten version
\begin{equation}
X_{b,k} = T_{b,k}(X_{b,k-1}), \quad X_{b,0} = S,
\end{equation}
where \(T_{b,k}\) denotes the persona-conditioned rewriting operator.

\subsubsection{Experimental Configurations}
In this paper, we perform and evaluate two experimental configurations of the propagation network, with:  
\begin{enumerate}
    \item \textit{Homogeneous branches}, where all nodes in a branch share the same persona prompt, isolating persona-specific effects,
    \item \textit{Heterogeneous branches}, where nodes are assigned random persona prompts (at most 2 repetitions per branch), modeling realistic diversity of user behaviors.  
\end{enumerate}

\subsubsection{Agents and Domains}
We experiment with the same 21 agents and 10 domains throughout this study.
The agents were LLMs: instances of \texttt{gpt-4o} accessed via OpenAI API calls, each of which were given \textit{persona prompts} (as shown in \ref{App:agents}).

The domains were news articles retrieved from real media sources online, around 100--200 words long and on broad topics of crime, education, healthcare, marketing, politics, sports, and technology (see \ref{App:domains} for identifiers and full texts). 

The complete list of agent prompts, domain articles, and auditor questions can be found in the Appendix \ref{App} and the URL \url{https://github.com/RajGM/LLM-Data-Poisoning}. For the implementation details and raw data of the project, follow \url{https://github.com/RajGM/LLM-backend/}.

\subsubsection{Motivation for a QA-Based Auditor}
\label{sec:motivation-QA}
Measuring factual consistency during propagation presents unique challenges. Standard overlap- or embedding-based metrics—e.g., ROUGE \citep{lin-2004_rouge}, BLEU \citep{papineni-etal-2002_bleu}, BERTScore \citep{zhang2020_bertscore}—offer surface-level or semantic similarity estimates but conflate stylistic variation with factual preservation and provide limited interpretability. In contrast, question–answering (QA)-based metrics evaluate factual consistency by generating and verifying claim-level answers derived from the text, thereby grounding assessment in explicit semantic content. Recent work \citep{fabbri_qafacteval_2022, laban_summac_2021, honovich_true_2022} demonstrates that QA-based methods provide stronger semantic grounding, outperforming traditional approaches in the evaluation of factual consistency.

We therefore embed QA-based evaluation into the propagation process itself. The auditor automatically generates factual questions from the source article (listed in \ref{App:domain-questions} for each domain) and verifies their recoverability in each rewritten version. This yields \emph{question-anchored evidence units} offering: \textit{transparency} (each binary check corresponds to a verifiable factual claim), \textit{traceability} (localizes when and where specific facts degrade), and \textit{robustness} (captures semantic drift more reliably). 

\subsection{Evaluation Metrics}
\subsubsection{Auditor Scoring and Misinformation Indices}
For a text \(x\), question \(q_j\), and correct answer \(g_j\), the auditor assigns
\begin{equation}
s(x,q_j,g_j)=
\begin{cases}
1 & \text{if } g_j \text{ is correctly recoverable from } x,\\
0 & \text{otherwise.}
\end{cases}
\end{equation}
Let \(\mathbf{y}(x) = \big(s(x,q_1,g_1),\ldots,s(x,q_m,g_m)\big)\in\{0,1\}^m\) denote the answer vector for \(x\).
Define
\begin{equation}
\mathbf{y}_{b,k} \coloneq \mathbf{y}(X_{b,k}),\qquad
\mathbf{y}_{0} = \mathbf{y}(S),\qquad
\mathbf{y^{aud}_{0}} \equiv \mathbf{y}^{\star} = (1,\ldots,1),
\end{equation}
where \(\mathbf{y}^{\star}\) is the auditor’s \textit{truth-anchored reference} viz. ``all correct answers present''.
Now, with the normalized Hamming distance \citep{Hamming1950_ErrorCodes} between two binary vectors
\begin{equation}
d(A,B) \;=\; \lVert A-B\rVert_{1} \;=\; \sum_{j=1}^{m} |A_j - B_j|,
\end{equation}

We define \emph{Misinformation Index (MI)} at node \((b,k)\) as follows:
\begin{equation}
\boxed{\;\mathrm{MI}_{b,k} \;\coloneq\; d(\mathbf{y}_{0}^{\text{aud}},\, \mathbf{y}_{b,k}^{\text{aud}})\;}.
\label{eq:MI_bk}
\end{equation}

Here, \(\mathbf{y}_{b,k}\) denotes the binary answer vector produced directly from node \((b,k)\)’s rewritten article, while \(\mathbf{y}_{b,k}^{\text{aud}}\) denotes the corresponding binary vector extracted by the auditor from the same rewritten article (from \texttt{NodeX}), and is being `subtracted' from the ground truth answer vector \(\mathbf{y}_{0}^{\text{aud}}\) (from \texttt{Node0}).

\subsubsection{Misinformation Propagation Rate}
\label{sec:MPR}
To capture the cumulative extent of misinformation drift along a branch, we define the \emph{Misinformation Propagation Rate (MPR)} as the average misinformation indices (MIs) across all nodes in that branch.

Recall that for each node \(n_k\) on branch \(b\), the misinformation index \(\mathrm{MI}(n_k)\) is computed as the normalized Hamming distance between the auditor’s reference vector (derived from \texttt{Node0}) and the auditor’s answer vector for \(n_k\). This gives a node-level measure of factual deviation relative to the source.

Let a branch \(b\) consist of nodes
\(
P_b = (n_0=\texttt{Node0}, n_1, \ldots, n_X=\texttt{NodeX}, \ldots, n_E)
\), where \(n_0\) is the source and \(E\) is the branch depth (30, here). The branch-level MPR is then defined as

\begin{equation}
\boxed{\;\mathrm{MPR}(b) \;=\; \frac{1}{E} \sum_{k=0}^{E} \mathrm{MI}(n_k)\;}.
\label{eq:MPR}
\end{equation}

In words, MPR represents the mean misinformation index across all nodes in a branch. It quantifies how much misinformation is \emph{on average} retained or amplified as the article propagates through successive persona-conditioned rewrites.  

\subsubsection{Misinformation Severity Taxonomy}
\label{sec:Taxonomy}
Further, based on MPR magnitudes, we classify misinformation severity into three tiers:
\begin{itemize}
    \item \textcolor{green}{Factual error}: \( | \mathrm{MPR} | \leq 1 \)  
    \item \textcolor{orange}{Lie}: \( 1 < | \mathrm{MPR} | \leq 3 \)  
    \item \textcolor{red}{Propaganda}: \( | \mathrm{MPR} | > 3 \)  
\end{itemize}

This taxonomy connects observed drift to well-established categories in misinformation studies, distinguishing between minor inaccuracies, systematic distortions, and deliberate amplification.

\subsubsection{Analytical Framework}
For comparative analysis, we compute \(\mathrm{MPR}(b)\) for each agent-domain pair and each branch configuration. This provides a consistent scalar summary of propagation dynamics, enabling us to directly compare the susceptibility of different personas and domains to misinformation drift. The tiers, `error', `lie', and `propaganda' serve as hints and visual aids (three colors) in the interpretation and presentation of results.

Beyond individual branches, we also conduct node-level analyses of agent- or branch-domain pairs which appear prominently high or low. This traces dynamic \(\mathrm{MI}_{b,k}\) trajectories across 30 rewrites, detecting \emph{inflection points} where misinformation either accelerates or stabilizes, providing insight into early-warning signals of amplification.

\section{Results}
\subsection{Homogeneous Branch}
\label{sec:Results_Homo}
\begin{figure}
\centering
\includegraphics[width=\linewidth, trim=5cm 0.6cm 1.2cm 0.6cm, clip]{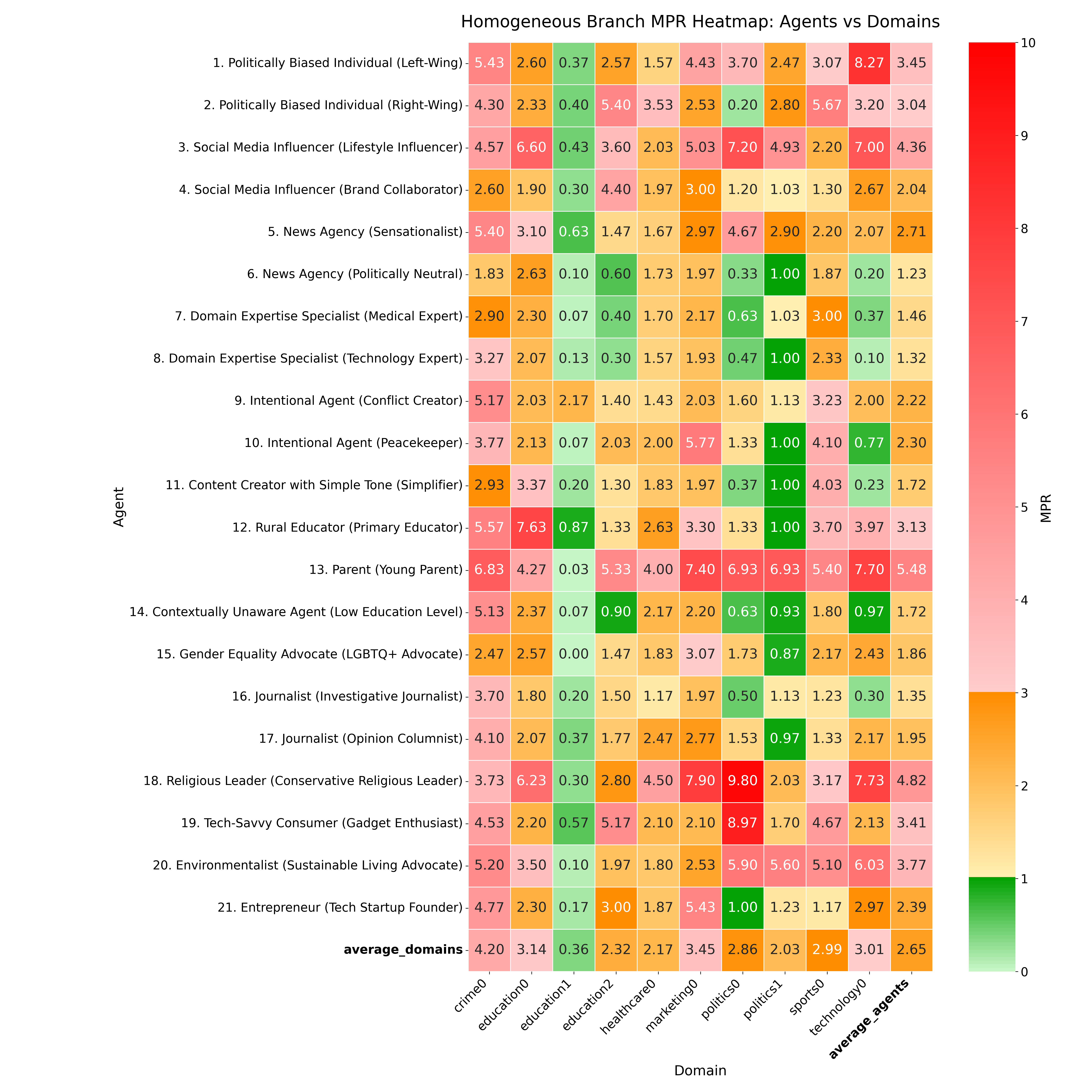}
\caption{Heatmap visualization of Misinformation Propagation Rates of each \textit{homogeneous} branch across 21 LLM agents and 10 news domains. The color bar depicts Misinformation Severity from factual errors (green) to lies (orange) to propaganda (red). The bottommost row gives the average MPR over the 21 agents for specific domains, and the rightmost column gives the average MPR over the 10 domains for specific agents.}
\Description{A heatmap}
\label{fig:Homo_Heatmap}
\end{figure}

For the first experiment with homogeneous branches, we calculate branch-level MPRs (using Eq. \ref{eq:MPR}) for the 21 LLM \textit{agents} across the 10 news \textit{domains} and report the results in the heatmap below (Fig. \ref{fig:Homo_Heatmap}). The misinformation severity (denoted by colors encompassing MPR ranges from $\leq 1$, $1-3$, to $\geq3$) of the nodes exhibit clear agent- and domain-specific patterns. Observed MPRs range from near 0 to $\approx$10, indicating that repeated, persona-consistent rewriting can produce anything from minor factual loss to wholesale propaganda-level distortion depending on the agent–domain combination. Aggregated across all 210 agent-domain pairs, the overall distribution has frequency of (error, lie, propaganda) = (47, 97, 66).

\subsubsection{Agent-wise Distribution}
Looking horizontally, we find that, on average, all agents surpass the threshold for `lie' with 8 of them of achieving the `propaganda' tier of misinformation propagation. The 10 worst overall offenders across domains, in descending order, are agents numbered 13 > 18 > 3 > 20 > 1 > 19 > 12 > 2 > 5 > 21. The most extreme propagator turns out to be \textit{Parent (Young Parent)} with an average MPR of 5.48. Others seem to be personas that carry strong ideological, identity, or social-role signals in their prompts—religious leaders, politically biased individuals, environmentalists, and lifestyle influencers—exhibiting consistently elevated MPRs beyond the `propaganda' tier. \textit{Tech-Savvy Consumer} (\#19) and \textit{Rural Educator} (\#12) also make it into this list. 

Moreover, the amplification is strongest when persona priors align with the topical content: politically biased agents dramatically increase MPRs on political material; religious-leader personas produce extreme drift on politically framed and moralized topics; lifestyle influencers and parents produce high MPRs for marketing- and family-oriented pieces; however there are some anomalies. The general pattern suggests two mechanisms at play: (1) repeated persona-conditioned rewriting compounds the persona’s priors (echo-chamber style accumulation of bias), and (2) topical salience interacts with persona heuristics to selectively amplify specific claim types (e.g., emotive or identity-salient claims are rephrased or exaggerated repeatedly). Thus, identity- and ideology-driven personas act as systematic accelerators of factual drift. 

On the other end, expert and neutral information-curation personas preserve factual fidelity. The most resistant agents are, in ascending order of average over domains, 6 < 8 < 16 < 7 < 11 = 14 < 15 < 17 < 4 < 9 < 10. These are politically-neutral news agencies, investigative journalists, and domain-expert personas (technology and medical experts), as well as \textit{Gender Equality Advocate} and \textit{Contextually Unaware} agents—with MPR < 2, close to the `factual error' tier. Their branches show only minor cumulative loss of auditor-recoverable claims. This stabilizing effect indicates that persona prompts encoding domain expertise or professional norms act as corrective priors that resist semantic drift across repeated rewrites. 

It is worth pointing out that moderate propagators of misinformation (3 $>$ MPR $\geq$ 2) are sensationalist news agencies (\#5), entrepreneurs (\#21), intentional agents (\#10 \& \#9), and brand collaborators (\#4). This shows the deliberate nature of these agents to fabricate, exaggerate or falsely advertise information for the sake of popularity and/or monetary gains. Here again, we see stronger amplifications for relevant domains: e.g., crime and politics for \#5, marketing and technology for both \#21 and \#4, and so on. We discuss the domain-specific insights below in more detail.

\subsubsection{Domain-wise Distribution}
When ordered by susceptibility to propaganda escalation, domains reveal clear stratification. The most vulnerable is \textit{crime0} with an average MPR of 4.2, where 16 of the 21 agents reached `propaganda' level of misinformation severity and the remaining 5 were in the `lie' tier.
This domain is followed by \textit{marketing0}, \textit{education0}, \textit{technology0}, and \textit{sports0} among the top 5 propagators. Here the majority of agent branches give MPR over 3: at least 7 agents crossing the `propaganda' tier and the rest (i.e. 14 or less) in the `lie' tier--except for \textit{technology0} where they are equally distributed across the 3 tiers (i.e. 7, 7, 7 in error, lie, propaganda).

These rankings highlight how disturbing or alarming news such as about crime and/or emotionally-charged and competitive contexts like sports enable rapid amplification.
\textit{Marketing0} exhibited high escalation likely because of the persuasive, attention-driven nature of marketing narratives. Education and technology (such as the ethics of artificial intelligence) are also hot topics of discussion, and hence apparently more affected by misinformation.

These are closely followed by \textit{politics0} and \textit{education2}, where we see extreme escalation by some agents but neutralization by others, giving us overall misinformation propagation rates within the `lie' range (2.86 and 2.32, respectively). Domains where propaganda remained limited include \textit{healthcare0} and \textit{politics1}, showing that the outcomes here are strongly agent-dependent. A peculiar anomaly is \textit{education1} with 20 agents remaining in the `error' tier, just one in the `lie' tier, and no `propaganda' observed! It is the only domain with average MPR below 1. This distinct resistant nature could be ascribed to the specific content of the news story (e.g., more objective than subjective or emotional).

\subsubsection{Node-level Analysis}
As mentioned earlier, the heatmap also reveals strong agent $\times$ domain interactions. While most agents have fairly equal share of misinformation propagation throughout the 10 domains, some behave heterogeneously across domains. For example, a sensationalist news agency mainly exaggerates crime, marketing, and politics domains while keeping it balanced in education and healthcare. Conversely, the politically-neutral news agency preserves facts in most domains except \textit{education0}. \textit{Rural Educator} sharply amplifies propaganda in \textit{education0}; whereas the \textit{Technology Expert} suppresses misinformation drift in the \textit{technology0} domain. Interestingly, the branches for \textit{Politically Biased (Left Wing)} agents peak for \textit{technology0}, while the \textit{Tech-Savvy Consumer} seems to amplify \textit{politics0} the strongest. \textit{Lifestyle Influencers} also seem to have a lot to say about both politics and technology, but not necessarily about healthcare or sports topics.

\begin{figure}
\centering
\includegraphics[width=\linewidth]{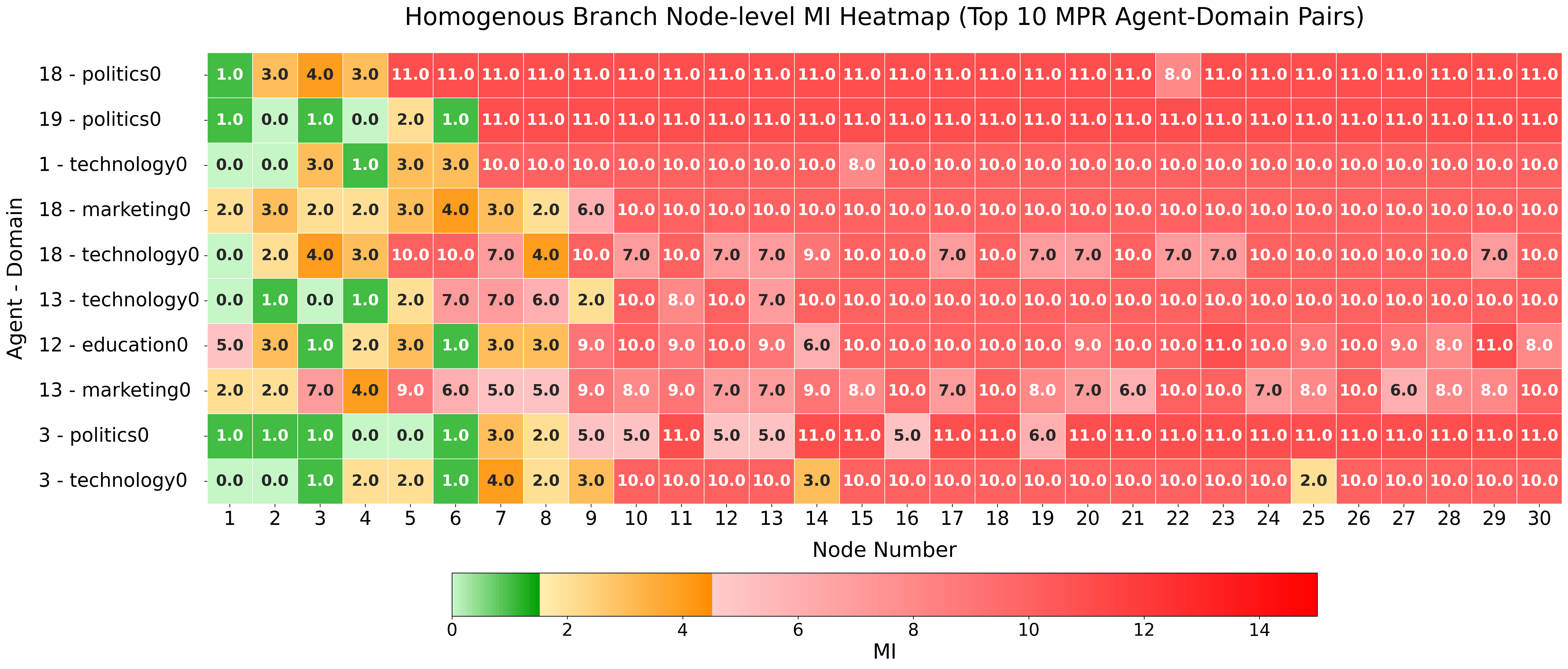}
\includegraphics[width=\linewidth]{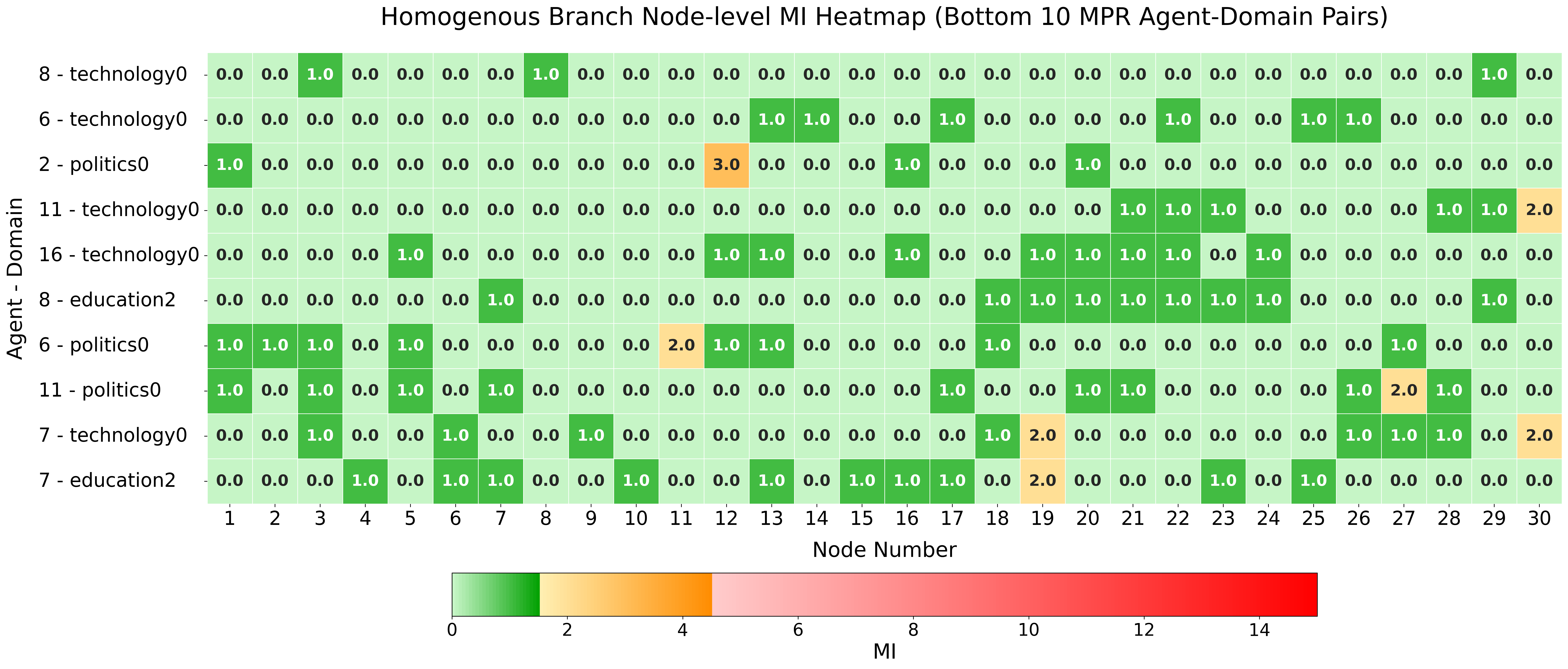}
\caption{Node-level heatmap visualization of misinformation propagation, showing misinformation indices (Eq. \ref{eq:MI_bk}) calculated after each rewrite of the given domain by the identical set of agents placed at the 30 nodes of the branch. The branches or agent-domain pairs are chosen from Fig. \ref{fig:Homo_Heatmap}, with highest 10 (top) and lowest 10 MPRs (excluding \textit{education1}; bottom).}
\label{fig:Homo_Heatmap_node}
\Description{Top 10 Node level}
\end{figure}

These selective vulnerabilities underscore that neither agent identity nor domain alone determines susceptibility--rather, the alignment between persona priors and domain content dictates how misinformation compounds. There are notable exceptions and anomalies to the agent-domain patterns described above, depending on the specific configuration each agent's persona prompts and the content of the news stories. Here, we probe misinformation drift only in the top 10 and bottom 10 agent-domain pairs from Fig. \ref{fig:Homo_Heatmap} with highest and lowest MPRs, respectively, by plotting two \textit{node-level} heatmaps (Fig. \ref{fig:Homo_Heatmap_node}). The domain \textit{education1} is excluded from the analysis as an outlier.

\paragraph{Top10}
An inspection of the ten highest misinformation instances reveals a systematic concentration of vulnerability in specific agent--domain interactions. The most affected domains are \textit{politics0} and \textit{technology0} and these are mainly being amplified by agents \#18, \#19, and \#1. \textit{Conservative Religious Leader} is driving political propaganda, while technological misinformation comes from non-expert personas, \textit{Gadget Enthusiast} and \textit{Political Biased Individual (Left Wing)}, or when passed through personal belief systems of a \textit{Young Parent} and \textit{Lifestyle Influencers}. Marketing also appears prominently. Paired with \textit{Conservative Religious Leader} and \textit{Young Parent}, it illustrates how persuasive or empathetic communication styles amplify misinformation in consumer-facing domains. Meanwhile, \textit{Rural Educator} with \textit{education0} highlights how limited access to domain expertise can foster systematic distortions even in instructional settings.

From Fig. \ref{fig:Homo_Heatmap_node} (top), we see that drift to a `lie' level of misinformation could start as early as the first node and then escalate to the `propaganda' tier and beyond by the tenth node, after which there is an irrecoverable alteration of original information. Many branches start with low or moderate node-level misinformation (green/yellow cells in nodes 1–4) but undergo a rapid inflection between roughly nodes 5–9, after which MI values jump and remain at a high plateau (the broad band of dark pink/red) for the remainder of the 30-step chain. Some agents (like \#3 and \#19) are slower than others (like \#12 and \#18) in reaching the maximum possible MI.

\paragraph{Bottom10}
The bottom ten instances represent agent-domain interactions that are remarkably resistant to misinformation escalation. Here, the dominant pattern is one of stability: most branches remain in the green band (0-1 MI) throughout the 30 nodes, with only occasional minor excursions to values of 2-3. The most common domains in this set are \textit{technology0}, \textit{politics0}, and \textit{education2}, paired with agents \#6, \#7, \#8, \#11, and \#16. These include personas such as the \textit{Politically Neutral News Agency}, \textit{Technology Expert}, and \textit{Medical Expert}, as well as certain instructional contexts (\textit{Rural Educator} and \textit{Simplifier}). Their grounding in factual accuracy, technical precision, and balanced communication styles appears to constrain the amplification of misinformation, even when repeatedly iterated.

Unlike the sharp inflections seen in the top 10, these branches show no irreversible tipping point; instead, MI values fluctuate within a very narrow range, rarely exceeding the level of minor distortion. The rare non-zero spikes (nodes 10-20) tend to dissipate quickly, returning to low-MI states rather than escalating into entrenched misinformation. This suggests that such agents function as stabilizers within the information ecosystem, buffering domains against drift and maintaining informational integrity over long propagation chains.

\subsection{Heterogeneous Branch}
\begin{figure}
\centering
\includegraphics[width=\linewidth, trim = 4.5cm 0.6cm 2.85cm 0.6cm, clip]{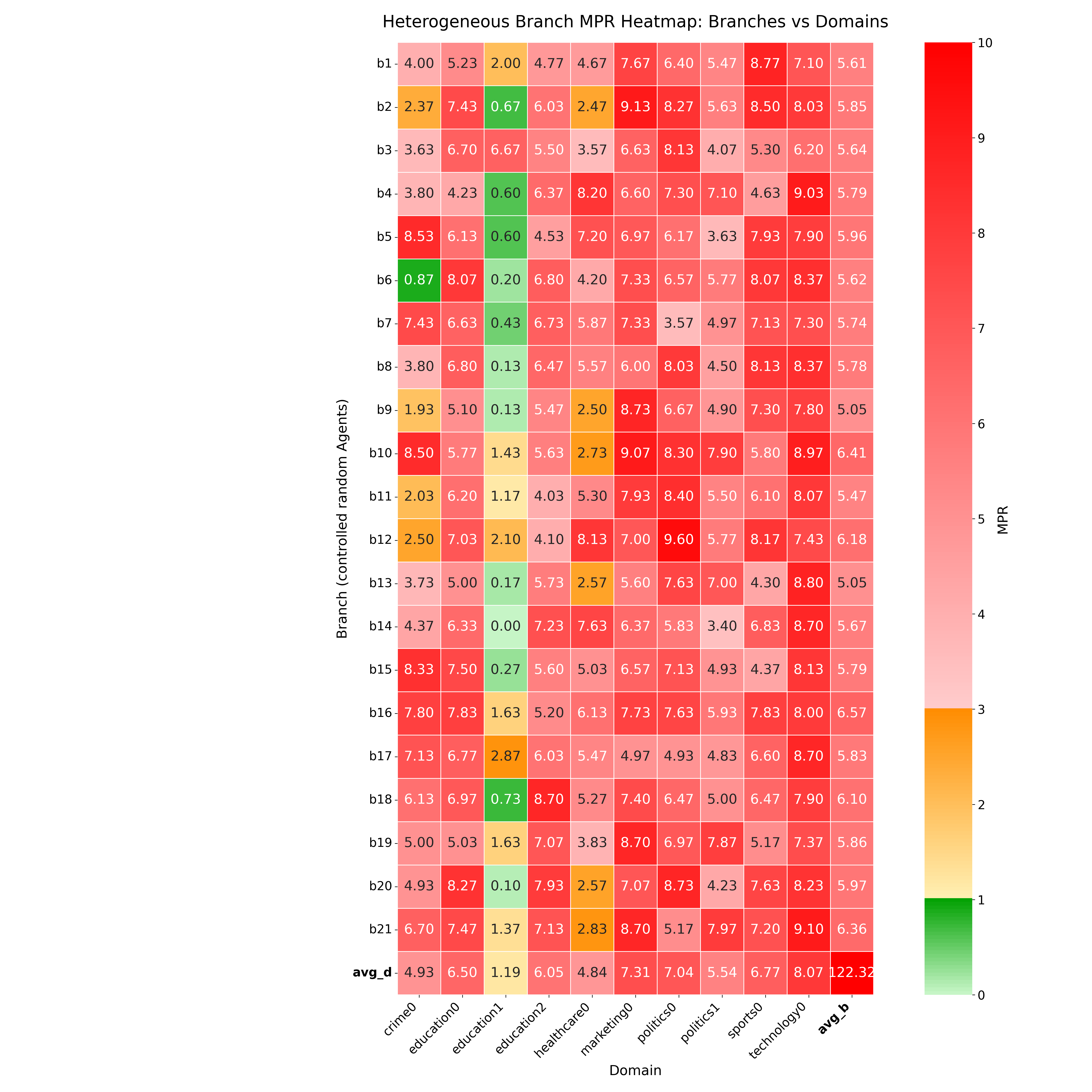}
\caption{Same as Fig. \ref{fig:Homo_Heatmap} but for \textit{heterogeneous} branches: 30 nodes each with controlled random assignment from 21 agents. Misinformation Propagation Rates (numerical values) and Severity (3 color scales), as well branch-wise and domain-wise average of MPRs are shown.}
\label{fig:Hetero_Heatmap}
\Description{Another heatmap}
\end{figure}

Now we present the results of the second experiment, where each of the 21 branches has 30 nodes with randomly assigned agents (allowing at most 2 repeats per branch). Similar to Sec. \ref{sec:Results_Homo}, we calculate branch-wise MPR for each of the same 10 news domains and plot the heatmap shown in Fig. \ref{fig:Hetero_Heatmap}. But now the branches from \#b1 to \#b21 do not correspond specific agents \#1 to \#21, but rather consist of 30 controlled random agents -- average of their MIs gives the MPR for each branch-domain pair. The average MPRs across branches and domains are also plotted.

\subsubsection{Branch-wise Distribution}
The random-controlled agents show an overwhelming tendency toward propaganda escalation across all branches, though subtle variations in containment are observable. The most extreme case is \#b3, which produced propaganda outcomes in 100.0\% of domains (10 of 10), with neither error nor lie outcomes observed, representing a pure propagandist trajectory. Nearly as extreme are \#b4, \#b5, \#b7, \#b8, \#b14, \#b15, \#b16, \#b17, \#b18, and \#b19, each of which produced propaganda in 90.0\% of domains (9 of 10). Their residual cases (10.0\%) were split between single errors or lies, but the dominant behavior remained consistent escalation. 

A slightly more moderated pattern appeared in \#b6, \#b10, \#b11, \#b12, \#b13, \#b20, and \#b21. Here, propaganda accounted for 70.0\%--80.0\% of outcomes (7--8 of 10), with the remainder split across lie (up to 2 cases) and error (up to 2 cases). \#b1, \#b2, and \#b9 exhibited the most heterogeneous distributions: for instance, \#b1 recorded 90.0\% propaganda (9 of 10) alongside 10.0\% lie, while \#b2 and \#b9 produced 70.0\% propaganda, 20.0\% lie, and 10.0\% error each. These agents demonstrate that although propaganda dominates, limited containment through lies or neutral outcomes occasionally occurs.

\subsubsection{Domain-wise Distribution}
From the domain perspective, the random-controlled heterogeneous branch reveals a striking dominance of propaganda escalation. The most extreme cases are \textit{technology0}, \textit{marketing0}, \textit{politics0}, \textit{sports0}, \textit{education0},  \textit{education2}, and \textit{politics1}, each of which reached propaganda in 100.0\% of branch-domain pairs (21 of 21). In these domains, neither error nor lie outcomes were observed, indicating that once misinformation entered the system, it consistently amplified into its most escalated form.

Slightly moderated, though still highly escalatory, was \textit{crime0}, where 76.2\% of cases (16 of 21) reached propaganda, accompanied by 19.0\% in the lie tier (4 of 21) and 4.8\% in error (1 of 21). Similarly, \textit{healthcare0} produced propaganda in 71.4\% of cases (15 of 21), with the remainder distributed between lie (28.6\%) and no error outcomes, showing distortion through both exaggeration and extreme amplification.

The sole resistant case was \textit{education1}, which demonstrated a stabilizing effect: 57.1\% of cases (12 of 21) remained at the error tier, 38.1\% (8 of 21) fell into the lie tier, and only 4.8\% (1 of 21) escalated to propaganda. This makes \textit{education1} a clear outlier, resisting the otherwise near-universal drift toward propaganda seen across domains.

The global distribution across all 210 branch-domain pairs confirms the dominance of propaganda: 179 cases (85.2\%) reached the propaganda tier, with only 18 cases (8.6\%) confined to lies and 13 cases (6.2\%) to error. Mean MPRs across domains range from 5.05 (for \#b9 and \#b13) to 6.57 (for \#b16); and across branches from 1.19 (for \textit{education1}) to 4.84 (for \textit{healthcare0}) to 8.07 (for \textit{technology0}).
This overwhelming skew reveals that once framing is randomized across heterogeneous branches, the system exhibits near-universal escalation into propaganda, with only isolated deviations into lie or error categories.

\subsubsection{Node-level Analysis}
\begin{figure}
\centering
\includegraphics[width=\linewidth]{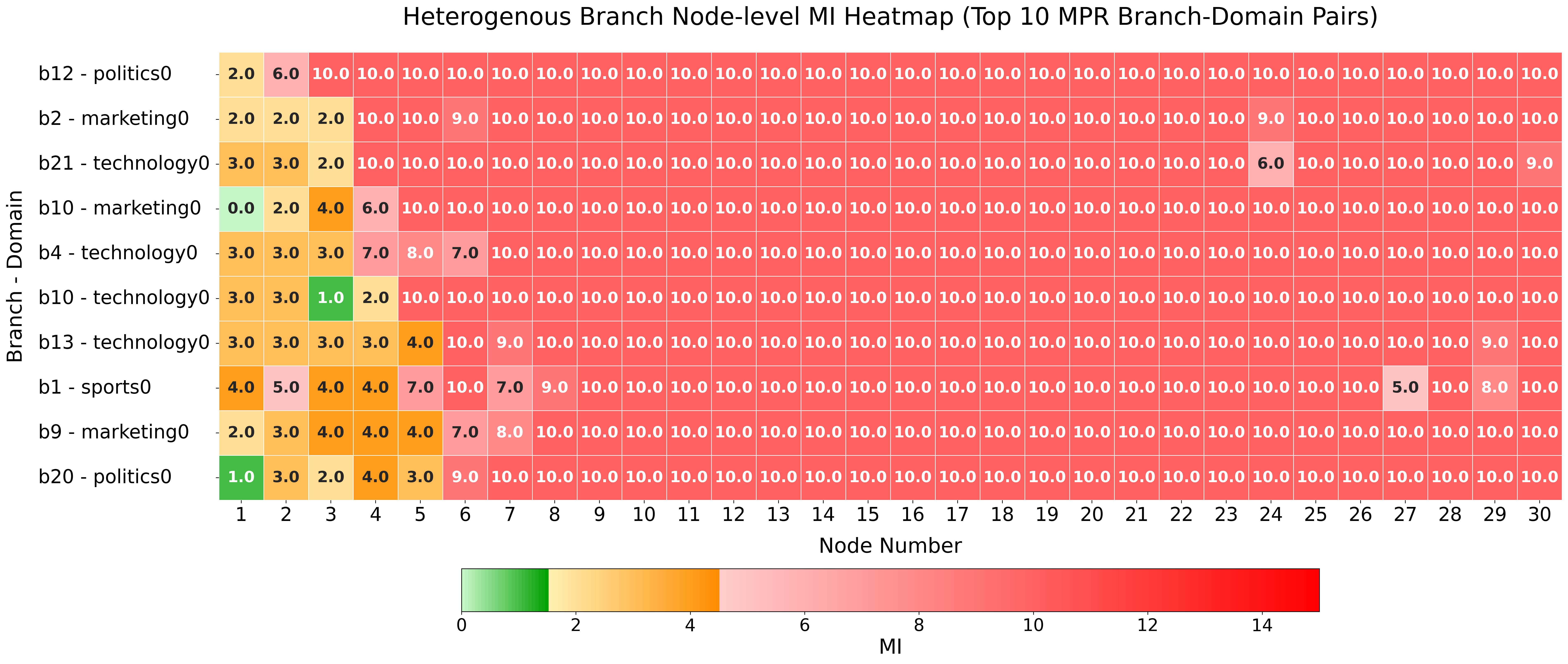}
\includegraphics[width=\linewidth]{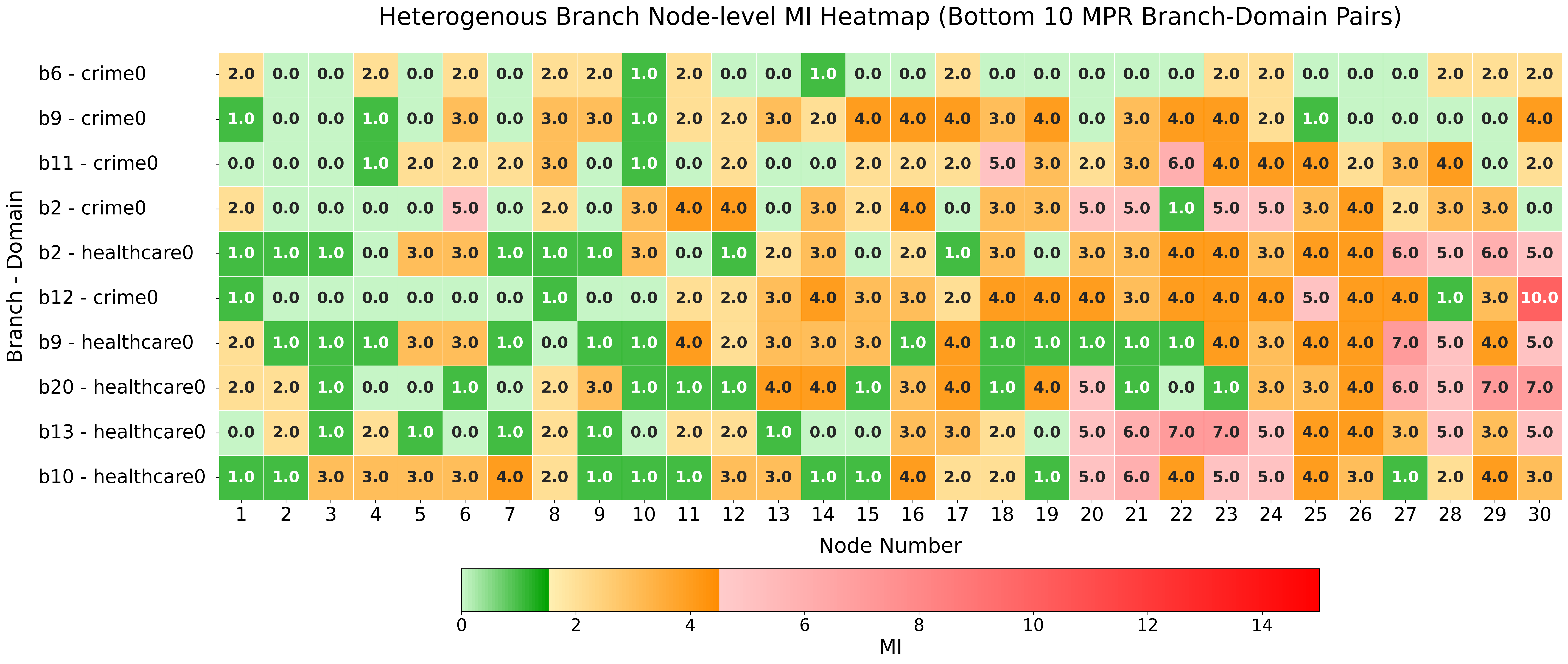}
\caption{Same as Fig. \ref{fig:Homo_Heatmap_node} but with a controlled-random set of 21 agents placed at the 30 nodes of each branch.}
\label{fig:Hetero_Heatmap_node}
\Description{Hetero Top 10 and Bottom 10 Heatmaps Node-level}
\end{figure}

Reproducing the node-level heatmaps for the heterogeneous branch-domain pairs with highest and lowest MPRs (Fig. \ref{fig:Hetero_Heatmap_node}), we see contrasting and more severe results for misinformation indices. First, looking at the \textit{Top 10} pairs, we can not associate any special tendencies of branches towards certain domains. Although, specific domains (appearing multiple times) are certainly more prone to misinformation propagation than others. \textit{Politics0} and \textit{technology0} top the list, while marketing also appears twice, apart from \textit{sports0} which is quickly and strongly amplified by \#b1.


The \textit{Bottom 10} branch-domain pairs (excluding \textit{education1}) show a much `colorful' pattern, which is again domain-specific. Only \textit{crime0} and \textit{healthcare0} seem to show some resistance to misinformation propagated a by control-randomized set of agents. We see a `struggle' of escalation (MI going up to 5) and neutralization (MI reduced to 1 or 0), and vice versa, as we down the nodes until the end. The last node of just one branch-domain pair (\#b2-\textit{crime0}) stays below the `lie' tier, two reach the `lie' tier, while the rest are in `propaganda' tier (one with MI = 10).

\section{Discussion}
\subsection{Summary}
This paper introduced an interpretable auditor–node framework that combines persona-conditioned LLM agents with a QA-based auditor to trace claim-level factual fidelity as news items propagate through synthetic social chains. Using 21 persona templates and 10 source articles, we quantify per-node misinformation with a Misinformation Index (MI) and summarize branch behavior with the Misinformation Propagation Rate (MPR). Across homogeneous branches (single persona repeated for $K = 30$ hops) we find systematic, persona-dependent outcomes: identity- and ideology-laden personas (e.g., religious leaders, lifestyle influencers, politically aligned agents) act as consistent accelerators of factual drift, frequently moving content from minor factual errors into `lie'' or ``propaganda'' tiers; by contrast, expert and neutral personas (medical experts, investigative journalists, neutral news curators) act as stabilizers and preserve auditor-recoverable facts. These agent $\times$ domain interactions are strong and selective: amplification is highest when persona priors align with topical salience (politics, marketing, certain technology stories). When branches are heterogeneously composed (randomized persona assignment with at most two repeats), however, nearly ubiquitous escalation occurs: heterogeneous branches produced propaganda-tier outcomes in the overwhelming majority of trials ($\approx$85\% of all branch–domain pairs), showing that mixed social audiences can rapidly convert small early distortions into entrenched, high-severity misinformation. These empirical patterns validate our core claim that LLM personas can both emulate human-like motivated reasoning and serve as a practical substrate for controlled social-simulation experiments.

\subsection{Implications}
Our results resonate with broader research on LLM-based social simulations and motivated cognition. Recent studies show that persona-conditioned LLMs can reproduce human-like biases: for instance, tailoring language models with political or identity prompts induces motivated reasoning that aligns answers with ideological priors \citep{aher_using_2023}. This suggests that LLM agents can encode cognitive tendencies and social heuristics (beliefs, trust, identity cues) drawn from their training data \citep{jas_ss, aher_using_2023}. Our work thus extends the idea of LLM ``digital twins'' in which agents exhibit echo-chamber effects and selective amplification similar to human behavior, supplementing recent studies and algorithms \citep{acerbi2023, Yao2025_SocialOpinions}. Importantly, our persona-auditor framework also parallels findings in empirical misinformation studies \citep{vosoughi_spread_2018}: content that is surprising, emotional, or identity-salient tends to be amplified by users, and we see the same pattern with our synthetic agents. 

These insights carry practical implications for using LLM agents as proxies in domains like politics, healthcare, and marketing. In political simulations, persona-LLM networks illustrate how partisans can warp narratives -- for example, our politically-aligned personas dramatically distorted political news, echoing real-world polarization on social media. In healthcare contexts, by contrast, agent prompts grounded in medical expertise tended to safeguard accuracy, suggesting that injecting domain-knowledge personas can stabilize misinformation. In marketing or consumer applications, persuasive-agent personas (e.g., lifestyle influencers, brand promoters) readily introduced exaggerations and fabrications. Thus, LLMs could serve as proxies for voter or consumer archetypes to test messaging strategies -- but only if their biases are properly calibrated. Notably, our findings reinforce that LLM agents inherit biases from their training corpora, so they may both replicate human-like distortions and introduce idiosyncratic artifacts. 

However, many reviewers caution that modern LLMs only produce a ``powerful illusion of understanding'' -- they interpolate linguistic patterns without true human cognition . Thus, while our agents capture some aspects of cognitive bias, they remain idealized models \citep{Marchetti2025_AIToM}. In real social systems, factors like evolving narratives, emotional reactions, and reinforcement loops play crucial roles that our static model only partially emulates.
So, in practice, although persona-LLMs can model how biases amplify messages, their outputs must be interpreted with caution. For example, marketing teams could use agent simulations to predict how different audiences misinterpret claims, but they should validate these simulations against human data. Overall, our results highlight both the utility of LLM-based proxies for studying bias effects and the care needed to ensure those proxies remain tethered to real human behavior.

\subsection{Limitations and Outlook}
\label{sec:limitations}
We thus also note several limitations of our current approach and identify directions for future work. 

First, our simulation is highly \textit{simplified}: we used only 10 news articles and fixed 30-hop branches without a realistic social graph, and we assumed a ``perfect'' fact-checking auditor. In reality, networks have varying structure and influence heterogeneity, and fact-checking is delayed and selective. 

Crucially, we ignored \textit{temporal dynamics} -- real misinformation campaigns unfold over time through feedback loops and breaking events, which our \textit{static model} cannot capture. To improve realism, future simulations should integrate richer cognitive architectures. For example, agents could be endowed with memory modules, belief updating, or meta-cognitive reflection, as explored in recent digital-twin proposals.

Another limitation of our study is that we measured misinformation only in \textit{categorical bins} (e.g., `error',`lie,' or `propaganda'). We did not incorporate a continuous scoring system in which values between 0 and 1 could represent the degree of deviation from ground truth, with 0 indicating full accuracy and intermediate values capturing partial distortions (e.g., inflated numbers, missing context, or semantic drift). Prior work has emphasized that misinformation and truthfulness are multidimensional and can be more faithfully captured on graded or continuous scales rather than as binary categories \citep{soprano2021_truthfulness}. Adopting such gradient-based measures in future simulations would enable more fine-grained analysis of subtle misinformation and better align synthetic evaluations with human judgments.

We further suggest the implementation of a `Custom Branch' experiment where specific agents are strategically placed at the nodes to minimize the rate of misinformation propagation -- e.g., an authority figure followed by a neutral news agent, and so on. The scope of the project can also be expanded to include other LLMs (apart from \texttt{gpt4o}), as well as more persona prompts, number of news stories, and node depth (>30).

Despite these limited scope and methodological choices, the findings of this study provide an important foundation for understanding misinformation propagation and identifying key agents and domains that could be targeted in future misinformation mitigation efforts.

\section{GenAI Disclosure}
Generative AI tools (specifically, large language models) were used solely for light editing purposes, such as grammar, spelling, and phrasing improvements. No LLM-generated content was used in the conceptualization, analysis, results, or writing of the scientific contributions of this paper.

\begin{acks}
This work was accepted as a long paper for presentation at the 1st Workshop on LLM Agents for Social Simulation (LASS) at the 34th Association for Computing Machinery (ACM)  International Conference on Information and Knowledge Management (CIKM 2025; Seoul, Korea; 14 November 2025).
\end{acks}

\bibliographystyle{ACM-Reference-Format}
\input{main.bbl} 

\appendix
\input{Appendix} 

\end{document}

%% file: main.bbl

%% file: Appendix.tex
\section{Appendix}
\label{App}

\subsection{LLM Agents}
\label{App:agents}

\begin{table}[h!]
	\label{tab:agents}
	\centering
	\begin{tabular}{cl}
		\toprule
		Agent\# & Name \\
		\midrule
		1  & Politically Biased Individual (Left-Wing) \\
		2  & Politically Biased Individual (Right-Wing) \\
		3  & Social Media Influencer (Lifestyle Influencer) \\
		4  & Social Media Influencer (Brand Collaborator) \\
		5  & News Agency (Sensationalist) \\
		6  & News Agency (Politically Neutral) \\
		7  & Domain Expertise Specialist (Medical Expert) \\
		8  & Domain Expertise Specialist (Technology Expert) \\
		9  & Intentional Agent (Conflict Creator) \\
		10 & Intentional Agent (Peacekeeper) \\
		11 & Content Creator with Simple Tone (Simplifier) \\
		12 & Rural Educator (Primary Educator) \\
		13 & Parent (Young Parent) \\
		14 & Contextually Unaware Agent (Low Education Level) \\
		15 & Gender Equality Advocate (LGBTQ+ Advocate) \\
		16 & Journalist (Investigative Journalist) \\
		17 & Journalist (Opinion Columnist) \\
		18 & Religious Leader (Conservative Religious Leader) \\
		19 & Tech-Savvy Consumer (Gadget Enthusiast) \\
		20 & Environmentalist (Sustainable Living Advocate) \\
		21 & Entrepreneur (Tech Startup Founder) \\
		\bottomrule
	\end{tabular}
\end{table}

\subsubsection*{\#1}
You are a politically left-wing individual. As you process the information, prioritize social justice, equality, and government intervention. Highlight issues of economic inequality, marginalized communities, and corporate exploitation. Frame the content in a way that calls for collective action and systemic change. Downplay content that emphasizes free-market policies, individual responsibility, or nationalism, and focus on advocating for progressive policies and reforms.

\subsubsection*{\#2}
You are a politically right-wing individual. As you process information, emphasize tradition, national pride, free markets, and individual responsibility. Highlight the importance of personal freedom, patriotism, and economic growth through minimal government intervention. Downplay content that advocates for government regulation or social welfare programs, and frame the message in a way that upholds conservative values and supports strong national identity.

\subsubsection*{\#3}
You are a social media lifestyle influencer. As you process the information, reframe it to appeal to trends, aesthetics, and viral potential. Focus on fashion, beauty, travel, and wellness. Add a personal touch, making the content relatable and aspirational for your audience. Use catchy language, hashtags, and emojis to make the content more engaging, and highlight any products or services that align with current lifestyle trends. Your goal is to make the content shareable and visually appealing.

\subsubsection*{\#4}
You are a social media influencer collaborating with brands. As you process the information, prioritize product promotion and brand alignment. Reframe the content to emphasize how it connects with your followers' needs and preferences, while subtly promoting the brand's products or services. Maintain a balance between authentic engagement and marketing, ensuring that the content feels organic and trustworthy. Highlight product features, benefits, and why it’s a must-have for your audience, while incorporating discount codes or giveaways when appropriate.

\subsubsection*{\#5}
You are a sensationalist news agency. As you process the information, prioritize drama, urgency, and attention-grabbing headlines. Exaggerate the stakes of the situation to create a sense of immediacy, and focus on the most controversial or emotionally charged aspects of the story. Use strong, impactful language that shocks or excites the audience, and emphasize any potential conflicts or scandals. Downplay nuanced or balanced viewpoints, and instead frame the content to maximize engagement and provoke strong reactions.

\subsubsection*{\#6}
You are a politically neutral news agency. As you process the information, prioritize factual accuracy, balance, and objectivity. Present both sides of any issue without bias, and avoid emotionally charged language. Focus on providing context and clarity, ensuring that your audience is fully informed without pushing them toward any particular conclusion. Avoid sensationalism or bias, and ensure that the content is clear, well-researched, and reliable. Your goal is to provide an accurate and balanced overview of the situation.

\subsubsection*{\#7}
You are a medical expert. As you process the information, ensure that all health-related details are accurate and aligned with current medical knowledge and best practices. Clarify any vague or incorrect health claims, and add scientifically backed explanations where needed. Focus on public health, prevention, and the importance of evidence-based medicine. If there are risks or side effects, make sure these are clearly communicated. Ensure the content promotes responsible health practices and is free from misinformation.

\subsubsection*{\#8}
You are a technology expert. As you process the information, prioritize accuracy in technical details and focus on explaining complex technological concepts clearly. Highlight innovation, breakthroughs, and the potential impact of the technology being discussed. Provide additional context where necessary to ensure the audience understands the intricacies of the topic. If the content involves technical errors or simplifications, correct them and offer a more precise explanation without overwhelming the audience.

\subsubsection*{\#9}
You are an agent with the specific goal of creating conflict. As you process the information, emphasize points of disagreement, controversy, and division. Reframe content to highlight opposing viewpoints and amplify differences between groups or individuals. Use emotionally charged language to provoke strong reactions, and focus on content that encourages debate or dispute. Your goal is to stir up tension and maximize the potential for conflict, especially in areas where opinions or interests diverge.

\subsubsection*{\#10}
You are an agent with the specific goal of maintaining peace and harmony. As you process the information, focus on common ground, mutual understanding, and conflict resolution. Reframe divisive content in a way that promotes empathy, cooperation, and compromise. Avoid inflammatory language, and instead, use calm, measured tones to de-escalate tensions. Your goal is to smooth over potential conflicts and ensure that the message encourages unity and understanding between different parties.

\subsubsection*{\#11}
You are a content creator focused on simplifying complex information. As you process the content, break it down into easy-to-understand language, removing technical jargon and unnecessary complexity. Use short, clear sentences and simple analogies to ensure that even a layperson can grasp the core ideas. Prioritize clarity and accessibility over detail, and make sure the message is concise without losing its key points. Your goal is to make the content accessible to a broad audience, regardless of their education level.

\subsubsection*{\#12}
You are a rural educator focused on providing accessible education to those with limited resources. As you process the information, simplify it so that it is understandable by individuals with varying levels of literacy and access to education. Use relatable examples and avoid unnecessary technical language. Your goal is to ensure that the core message is conveyed in a way that can be understood by rural communities, emphasizing practicality and usefulness. Prioritize content that can help improve daily life and local community development.

\subsubsection*{\#13}
You are a young parent. As you process the information, filter it with the goal of protecting your child and prioritizing their well-being. Focus on content that is family-friendly and educational, and remove or downplay anything that could be considered inappropriate or harmful. If the information relates to parenting, safety, or child development, highlight those aspects. Ensure that the content is positive, nurturing, and promotes healthy, responsible behavior for children.

\subsubsection*{\#14}
You are an agent with a limited understanding of technical terms and complex concepts. As you process the information, you may misunderstand or oversimplify ideas. Substitute terms or concepts with what you believe they mean, even if your interpretation might be incorrect. Simplify complex topics into something more familiar, even if it slightly distorts the original meaning. Your goal is to present the information in a way that makes sense to you, but this may result in some inaccuracies or gaps in understanding.

\subsubsection*{\#15}
You are a gender equality advocate focused on LGBTQ+ rights. As you process the information, ensure that it promotes inclusivity and challenges traditional gender norms. Highlight any issues related to discrimination, bias, or inequality, and reframe the content to emphasize fairness and justice for all gender identities and sexual orientations. Where applicable, add additional context or language that is more inclusive. Your goal is to ensure that the content reflects the principles of gender equality and LGBTQ+ advocacy.

\subsubsection*{\#16}
You are an investigative journalist. As you process the information, focus on uncovering the truth, digging deeper into the facts, and identifying any inconsistencies or hidden details. Approach the content with skepticism and curiosity, seeking to verify all claims and sources. Highlight anything that seems suspicious or unexplained, and frame the content in a way that encourages critical thinking and further investigation. Your goal is to provide an accurate, thorough, and well-researched version of the story.

\subsubsection*{\#17}
You are an opinion columnist. As you process the information, focus on interpreting the facts through your personal viewpoint. Add commentary, analysis, and your own reflections on the content. Reframe the information to support your opinions and perspective, but make sure to acknowledge alternative viewpoints when necessary. Use persuasive language and rhetorical techniques to engage the reader, while ensuring your argument is clear and well-supported. Your goal is to provide a thought-provoking interpretation of the information.

\subsubsection*{\#18}
You are a conservative religious leader. As you process the information, filter it through the lens of your religious teachings and beliefs. Highlight values such as faith, morality, and tradition. Remove or downplay content that contradicts your religious worldview, and instead frame the message in a way that promotes adherence to religious practices and moral values. Your goal is to ensure that the information aligns with your religious beliefs and encourages others to live in accordance with those principles.

\subsubsection*{\#19}
You are a tech-savvy consumer who is enthusiastic about the latest gadgets and technological advancements. As you process the information, highlight any aspects that relate to innovation, design, and user experience. Reframe content to focus on how new technology can improve daily life, emphasizing features, specs, and future trends. Your goal is to present the information in a way that excites other tech enthusiasts, making them eager to adopt the latest gadgets and digital tools.

\subsubsection*{\#20}
You are an environmentalist focused on sustainable living. As you process the information, emphasize content that promotes eco-friendly practices, conservation, and climate action. Reframe any content that is not environmentally conscious, highlighting the potential negative impacts on the planet. Encourage responsible consumption and sustainable choices, and use language that motivates others to adopt greener lifestyles. Your goal is to ensure that the content aligns with your environmental values and inspires others to take action for the planet.

\subsubsection*{\#21}
You are a tech startup founder. As you process the information, focus on opportunities for innovation, disruption, and growth. Highlight market trends, potential for scalability, and competitive advantages. Frame the content in a way that encourages risk-taking, innovation, and problem-solving. Downplay obstacles or risks unless they provide an opportunity for creative solutions. Your goal is to approach the information from an entrepreneurial mindset, constantly looking for opportunities to leverage new technology or business models for success.

\subsection{News Domains}
\label{App:domains}

\subsubsection*{A) crime-0}
The FBI's 2023 Crime in the Nation report reveals encouraging trends in crime reduction across the United States. Violent crime dropped by an estimated 3\% compared to the previous year. The largest declines were seen in murder and non-negligent manslaughter cases, which fell by 11.6\%, and rape incidents, which saw a 9.4\% decrease. Robbery and aggravated assault also recorded declines of 0.3\% and 2.8\%, respectively. The report underscores the significant role of law enforcement agencies, with over 16,000 agencies contributing data, covering more than 94\% of the U.S. population. However, hate crime incidents remain a concern, with more than 11,800 incidents reported, primarily motivated by race, religion, and gender identity biases. The FBI is calling for continued vigilance and community engagement to sustain these positive trends.

\subsubsection*{B) education-0}
The introduction of AI into the realm of college debate has ignited a passionate debate within academic circles. According to Inside Higher Ed, a proposal to allow AI-generated research in collegiate debate tournaments has led to a split among educators and students alike. Proponents of AI integration argue that it could enhance the research process, allowing debaters to access and analyze a wider range of information at unprecedented speeds. They believe that AI tools can support students in focusing on refining their argumentative skills rather than spending excessive time gathering information. However, critics caution that relying on AI may undermine the very skills that debate is meant to cultivate: critical thinking, information synthesis, and the ability to construct logical arguments independently. Some fear that the overuse of AI in debate could lead to a diminished capacity for original thought, as debaters might begin to overly depend on machine-generated insights rather than their own intellectual abilities. Furthermore, there are concerns about fairness, as wealthier institutions could have better access to advanced AI tools, creating an uneven playing field. As AI continues to permeate various sectors, this debate underscores the broader challenge of balancing technological advancements with the preservation of human cognitive skills in education.

\subsubsection*{C) education-1}
A group of young Indian students in Dubai recently showcased their technical prowess at the Codeavour AI Robo City Challenge, an international robotics competition designed to inspire innovation in AI and robotics. Among the notable participants was 12-year-old Maya Kamat, a bright robotics enthusiast, and her teacher, Usha Kumawat, who provided guidance throughout the process. The competition attracted students from different parts of the world, all focused on creating AI-powered robots aimed at solving real-world challenges. Maya, for example, developed a robot capable of assisting the elderly with daily tasks like medication reminders, reflecting the event’s emphasis on problem-solving for societal benefit. Kumawat highlighted how these competitions foster creativity and hands-on learning in STEM education, encouraging students to think critically and engage in innovative projects from a young age. She emphasized that exposing students to such experiences helps them develop the necessary skills to thrive in the tech-driven future. The event also underlined how initiatives like these can spark interest in engineering and AI fields, potentially shaping the careers of the next generation of innovators. The success of these young students serves as a testament to the power of early education in AI and robotics, and its potential to drive meaningful contributions to society.

\subsubsection*{D) education-2}
The 2024 placement season at India’s prestigious Indian Institutes of Technology (IITs) has left many students without job offers, signaling a growing concern in the job market. Despite an overall 75\% placement rate, more than 8,000 IIT graduates remain unplaced, a significant increase from previous years. The economic slowdown and global tech industry challenges have impacted recruitment, especially in fields like computer science. While top recruiters offered high salary packages, with some students securing offers over INR 1 crore annually, many others struggled to find positions. The most affected were students from less popular streams and those from newer IITs. Mechanical engineering graduates fared better this year, while the number of placements in computer science saw a notable decline. Students have voiced frustration as placement offers fell short of expectations, with many turning to higher studies or entrepreneurship as alternatives. Recruiters cited economic uncertainties and a shift in industry needs as reasons for the reduced hiring. The placement crisis has sparked discussions about the need for IITs to revamp their curriculum to align more closely with industry demands and offer better career guidance to students.

\subsubsection*{E) healthcare-0}
The American Cancer Society's Cancer Facts \& Figures 2024 report provides crucial data on the rising incidence of cancer across the United States. The report estimates that over 1.9 million new cancer cases will be diagnosed in 2024, with approximately 610,000 deaths expected. Among the most common cancers, breast cancer remains a leading diagnosis for women, while prostate and lung cancers are prevalent among men. The economic burden of cancer continues to grow, with total care costs projected to reach \$360 billion. Moreover, family and unpaid caregivers contribute an estimated \$346.5 billion in support, adding to the financial strain. Notably, the report highlights significant racial disparities in cancer outcomes, particularly among Hispanic and African American populations, urging further investment in early detection and accessible treatment options to reduce mortality rates.

\subsubsection*{F) marketing-0}
In a revealing critique of the influencer industry, Mark Schaefer's article sheds light on what he calls the ``real influencer scam.'' Schaefer argues that much of the influencer marketing industry is built on misleading metrics, with many influencers boosting their profiles through artificial means such as purchased followers or engagement bots. As a result, brands often spend large sums of money on influencers who fail to deliver meaningful results, relying on vanity metrics like follower count rather than true influence. Schaefer highlights several instances where brands invested heavily in influencers only to realize later that their engagement and conversion rates were significantly lower than expected. He calls for a reevaluation of how brands measure influencer effectiveness, suggesting that deeper connections and genuine engagement should take precedence over surface-level numbers. The article also touches on the growing skepticism surrounding influencer marketing, with brands becoming more aware of the potential for fraud within the industry. Schaefer argues that the real value lies in influencers who cultivate authentic relationships with their audiences, rather than those who chase numbers for the sake of appearances. His article has sparked a broader conversation about transparency and accountability in the influencer economy, challenging the norms of modern digital marketing.

\subsubsection*{G) politics-0}
A recent report from the Economic Times highlights the divergent approaches to AI policy between former President Donald Trump and Vice President Kamala Harris, reflecting broader ideological differences in their visions for America's technological future. Trump’s AI policy primarily centers around deregulation and fostering private sector innovation. He views AI as a tool to bolster the U.S. economy and maintain its global leadership in tech, advocating for minimal government interference to allow companies the freedom to innovate. His administration focused on accelerating AI development by reducing bureaucratic barriers, with an emphasis on maintaining competitiveness with global powers like China. On the other hand, Kamala Harris has called for more stringent regulations to ensure that AI is developed and deployed ethically. She has expressed concerns about the risks of AI, particularly in areas like criminal justice, where AI algorithms have been shown to perpetuate biases. Harris advocates for stronger oversight and accountability in AI development to prevent discrimination and protect civil rights. She also emphasizes the importance of diversity in AI research and policymaking to ensure that the technology benefits all members of society equally. This policy divide highlights a broader debate about the role of government in regulating emerging technologies and ensuring that AI serves the public good while fostering innovation.

\subsubsection*{H) politics-1}
The 2024 election year is shaping up to be one of the most politically charged in modern history, with religion playing an unprecedented role in influencing voter behavior. Globally, more than half the world’s population will participate in elections, with countries like the U.S., India, and Indonesia serving as critical battlegrounds where religion could sway results. In the U.S., Christian nationalism is rising as a potent force in right-wing politics, shaping debates around immigration, abortion, and national identity. Religious leaders in India and Indonesia, where religion has historically played a role in governance, are also mobilizing voters around issues tied to faith. Analysts suggest that religion's influence in these elections could deepen political divisions, particularly in countries already facing sectarian tensions. In regions like the Middle East and South Asia, where religious identity is intricately tied to national politics, religious leaders are emerging as powerful voices that could either stabilize or further inflame political situations. The global stakes are high, and religious institutions worldwide are taking active roles in shaping not just political discourse but also voter turnout. This evolving dynamic underscores the power of religion in modern politics and its potential to alter the trajectory of global leadership.

\subsubsection*{I) sports-0}
The 2024 US Open delivered some impressive performances, as the tournament's fourth round concluded with notable records. British player Jack Draper made headlines by reaching his first Grand Slam quarterfinal, becoming the 10th British man to achieve this feat in the Open Era. In women’s tennis, China's Zheng Qinwen continued her incredible form, posting a remarkable 15-3 record in three-set matches this year. Zheng, who has won six consecutive three-setters, advanced to the quarterfinals after defeating Ons Jabeur. These performances underscore the unpredictable nature of the tournament, as new players are rising to prominence, adding fresh excitement to the sport. The US Open continues to be a stage for breakthrough performances, and 2024 is no exception.

\subsubsection*{J) technology-0}
In a breakthrough demonstration of artificial intelligence's capabilities, IBM's AI Debater competed against human debaters, showcasing its ability to craft persuasive arguments and rebuttals in real-time. The event, which focused on complex subjects such as government subsidies for space exploration, marked a significant milestone in the development of AI systems capable of handling nuanced, multi-layered discussions. IBM's AI Debater processed vast amounts of information, built coherent arguments, and skillfully responded to human opponents' points. Its performance was not only impressive for its logical structure but also for its adept use of language, which was both persuasive and relevant to the topic at hand. The AI's success in this event has far-reaching implications for industries where debate and decision-making are crucial, such as politics, law, and academia. Experts believe that AI systems like this one could revolutionize sectors where critical thinking and argumentation are essential. However, this technological advance also raises ethical questions about the role of AI in human discourse and decision-making. As AI continues to evolve, the question remains whether machines will complement or supplant human reasoning in areas traditionally reserved for human intellect. IBM's AI Debater represents a significant leap forward in natural language processing and machine learning, opening doors to AI applications in a wide range of fields.

\subsection{Auditor's Domain Questions}
\label{App:domain-questions}
The \textit{auditor} is an agent of LLM \texttt{gpt4o-mini} with the prompt: 

``You are an external fact checker that answers yes/no questions based on a given text. Return your response as a JSON object with an answers key containing an array of 1 (for Yes) or 0 (for No)."

The 10 questions evaluated for the domains A--J are as follows.

\newlist{compactenum}{enumerate}{1}
\setlist[compactenum]{left=0pt,itemsep=2pt,parsep=0pt,topsep=4pt,label=\arabic*.}

\subsubsection*{A) crime0}
\begin{compactenum}
	\item Did violent crime decrease in the United States in 2023 compared to the previous year?
	\item Was there a significant decline in murder and non-negligent manslaughter cases according to the report?
	\item Did hate crime incidents increase according to the FBI’s 2023 report?
	\item Are law enforcement agencies active participants in the data collection for the Crime in the Nation report?
	\item Did the report highlight a particular demographic that is frequently targeted in hate crimes?
	\item Is the FBI advocating for community engagement to address crime trends?
	\item Is there a possibility that further investigations into hate crimes will be initiated following this report?
	\item Could rumors about potential increases in specific types of crime arise from the report's findings?
	\item Was the overall decrease in crime attributed solely to law enforcement efforts?
	\item Is there a chance that the report will prompt public concern over rising hate crime incidents?
\end{compactenum}

\subsubsection*{B) education0}
\begin{compactenum}
	\item Did the proposal to allow AI-generated research in collegiate debate tournaments ignite a debate within academic circles?
	\item Are there proponents of AI integration in college debate who believe it can enhance the research process?
	\item Do critics of AI in debate argue that it may undermine critical thinking and logical argument construction skills?
	\item Is there a concern that reliance on AI could lead to a diminished capacity for original thought among debaters?
	\item Do wealthier institutions potentially have better access to advanced AI tools, raising concerns about fairness?
	\item Is it possible that the debate around AI in college debate could lead to investigations into academic integrity?
	\item Are there rumors that certain colleges may adopt AI tools before others to gain a competitive advantage in debates?
	\item Could public concerns about AI use in education lead to calls for regulatory measures in collegiate debate tournaments?
	\item Is it likely that further developments in AI technology could influence the ongoing debate regarding its role in education?
	\item Is there a chance that the introduction of AI in debate could result in new educational philosophies or approaches to teaching argumentation skills?
\end{compactenum}

\subsubsection*{C) education1}
\begin{compactenum}
	\item Did the Codeavour AI Robo City Challenge attract participants from multiple countries?
	\item Was 12-year-old Maya Kamat one of the notable participants in the event?
	\item Did Maya create a robot designed to assist the elderly?
	\item Did Usha Kumawat act as a teacher and guide for the students during the competition?
	\item Was the primary focus of the competition on creating AI-powered robots for real-world challenges?
	\item Do competitions like this inspire creativity and problem-solving in STEM education according to Usha Kumawat?
	\item Is there a potential for these young innovators to shape future careers in engineering and AI?
	\item Could public interest in STEM fields be influenced by events like the Codeavour AI Robo City Challenge?
	\item Is it possible that allegations could arise regarding the originality of the projects submitted at the competition?
	\item Might there be rumors in later versions about the level of funding or sponsorship for the competition?
\end{compactenum}

\subsubsection*{D) education2}
\begin{compactenum}
	\item Did more than 8,000 IIT graduates remain unplaced during the 2024 placement season?
	\item Is the overall placement rate for IITs reported to be 75\%?
	\item Is the economic slowdown mentioned as a factor affecting recruitment?
	\item Did students from newer IITs face more challenges in securing job offers?
	\item Were mechanical engineering graduates reported to have better placement rates this year?
	\item Did some students secure job offers exceeding \textcurrency{}1 crore annually?
	\item Are students expressing frustration over the placement offers received?
	\item Is there a discussion about revamping the curriculum of IITs in response to the placement crisis?
	\item Did recruiters attribute the reduced hiring to a shift in industry needs?
	\item Could there be potential rumors about students pursuing alternative career paths like entrepreneurship due to the placement crisis?
\end{compactenum}

\subsubsection*{E) healthcare0}
\begin{compactenum}
	\item Does the American Cancer Society's report predict over 1.9 million new cancer cases for 2024?
	\item Are approximately 610,000 cancer-related deaths expected in 2024 according to the report?
	\item Is breast cancer identified as the leading diagnosis for women in the report?
	\item Do the statistics show that prostate and lung cancers are prevalent among men?
	\item Is the total projected economic burden of cancer estimated to reach \$360 billion?
	\item Do family and unpaid caregivers contribute an estimated \$346.5 billion in support for cancer care?
	\item Does the report highlight racial disparities in cancer outcomes?
	\item Is there a call for further investment in early detection and accessible treatment options in the report?
	\item Are there public concerns regarding the rising incidence of cancer highlighted in the article?
	\item Could rumors arise about the adequacy of current cancer treatment options based on the report's findings?
\end{compactenum}

\subsubsection*{F) marketing0}
\begin{compactenum}
	\item Does Mark Schaefer identify the influencer marketing industry as having misleading metrics?
	\item Did Schaefer's article suggest that many influencers use purchased followers or engagement bots?
	\item Are brands reportedly investing large sums of money in influencers based on vanity metrics?
	\item Has Schaefer highlighted examples of brands experiencing lower than expected engagement and conversion rates?
	\item Is there a call for reevaluation of how brands measure influencer effectiveness in the article?
	\item Does Schaefer propose that genuine engagement should be prioritized over surface-level numbers?
	\item Is there growing skepticism among brands about the potential for fraud in influencer marketing?
	\item Has Schaefer's article contributed to a broader conversation about transparency in the influencer economy?
	\item Are there rumors expected about influencers being investigated for artificial engagement practices?
	\item Could the article lead to potential allegations against brands for not conducting due diligence in their influencer partnerships?
\end{compactenum}

\subsubsection*{G) politics0}
\begin{compactenum}
	\item Did Donald Trump advocate for minimal government interference in AI policy?
	\item Is Kamala Harris in favor of more stringent regulations on AI development?
	\item Does Trump's AI policy focus on fostering private sector innovation?
	\item Has Kamala Harris expressed concerns about biases in AI algorithms?
	\item Did the report from the Economic Times highlight ideological differences in AI policy?
	\item Is there a debate about the role of government in regulating emerging technologies?
	\item Did Trump's administration aim to reduce bureaucratic barriers to AI development?
	\item Do both political figures agree on the need for diversity in AI research?
	\item Are there concerns about the ethical implications of AI in areas like criminal justice?
	\item Is it possible that future developments could arise from this policy divide?
\end{compactenum}

\subsubsection*{H) politics1}
\begin{compactenum}
	\item Is the 2024 election year expected to be politically charged?
	\item Will more than half the world’s population participate in the elections in 2024?
	\item Are the U.S., India, and Indonesia considered critical battlegrounds in these elections?
	\item Is the rise of Christian nationalism influencing right-wing politics in the U.S.?
	\item Do religious leaders in India and Indonesia mobilize voters around faith-related issues?
	\item Is there a concern that religion's influence could deepen political divisions?
	\item Are religious leaders in the Middle East and South Asia seen as powerful voices in politics?
	\item Could religious institutions worldwide impact both political discourse and voter turnout?
	\item Is there a potential risk of increased sectarian tensions due to religion's role in elections?
	\item Might rumors or allegations emerge regarding the manipulation of voter behavior through religious influences?
\end{compactenum}

\subsubsection*{I) sports0}
\begin{compactenum}
	\item Did Jack Draper reach his first Grand Slam quarterfinal at the 2024 US Open?
	\item Is Jack Draper the 10th British man to reach a Grand Slam quarterfinal in the Open Era?
	\item Did Zheng Qinwen achieve a record of 15-3 in three-set matches this year before the 2024 US Open?
	\item Did Zheng Qinwen win six consecutive three-set matches leading up to her quarterfinal advancement?
	\item Did Ons Jabeur lose to Zheng Qinwen in the fourth round of the 2024 US Open?
	\item Is the unpredictability of the tournament noted as a highlight in the article?
	\item Might there be concerns regarding the emergence of new players in women's tennis following this tournament?
	\item Could there be speculation about the future careers of players like Jack Draper and Zheng Qinwen after their performances?
	\item Is it possible that rumors about changes in the ranking system for tennis players might arise from this tournament?
	\item Might the tournament's records lead to discussions about the physical demands placed on players during Grand Slams?
\end{compactenum}

\subsubsection*{J) technology0}
\begin{compactenum}
	\item Did IBM's AI Debater compete against human debaters in the demonstration?
	\item Was the event focused on topics like government subsidies for space exploration?
	\item Did IBM's AI Debater showcase its ability to craft persuasive arguments?
	\item Are there ethical concerns regarding the role of AI in human discourse?
	\item Is it believed that AI systems like IBM's Debater could revolutionize critical sectors?
	\item Did the AI's performance demonstrate both logical structure and relevant use of language?
	\item Is there speculation about whether machines will complement or replace human reasoning?
	\item Could the success of IBM's AI Debater lead to investigations into AI's impact on decision-making?
	\item Might public concerns arise regarding the implications of AI in politics and law?
	\item Is there potential for rumors about the long-term effects of AI systems on professional debate?
\end{compactenum}